\documentclass[letterpaper]{article}
\usepackage{aaai}
\usepackage{times}
\usepackage{helvet}
\usepackage{courier}
\usepackage{graphicx} 
\usepackage{pifont}
\usepackage{amsmath}
\usepackage{caption}
\usepackage{float}
\usepackage{placeins}
\usepackage{booktabs}
\usepackage{enumitem}
\setlist[enumerate,1]{label=(\arabic*), leftmargin=2em}
 
\begin{document}
%
\title{Multi Agents Semantic Emotion Aligned Music to Image Generation \\ with Music Derived Captions}
\author{JunChang Shi,Gang Li\textsuperscript{*}\\
School of Computer \& Communication Engineering,\\University of Science and Technology Beijing,China\\
\texttt{M202320906@xs.ustb.edu.cn, ligang6867@ustb.edu.cn}\\
\textsuperscript{*}Corresponding author: ligang6867@ustb.edu.cn
}

\maketitle
\begin{abstract}
\begin{quote}
When you listen to music, what images appear in your mind? 
This work helps you generate them. 
We address the scarcity of direct music-to-image generation by proposing a multi-agent semantic–emotion aligned pipeline. 
A captioning component first produces structured music-derived semantics; cooperating agents then extract style (e.g., genre, tempo) and affect attributes to form a joint conditioning representation for the image synthesis module. 
To quantify and reinforce emotional coherence, we fine-tune a CLIP ViT backbone \cite{dosovitskiy2021vit,radford2021clip} with a visual–emotion regression head trained on the Emotic dataset, which provides 26 discrete emotion categories together with continuous valence–arousal–dominance annotations. 
Empirically, the performance of emotion prediction in our image-emotion regression head and music caption module is comparable to prior image emotion recognition \cite{zhao2020emotion} and unified music emotion recognition \cite{kang2025unified}
Empirical results show that this regression-augmented multiagent design yields higher semantic–emotion consistency between generated images and source music than caption-only baselines, demonstrating an effective step toward cognitively plausible visual synthesis from music-derived captions and affective attributes.
\end{quote}
\end{abstract}

\section{Introduction}
Music and vision are deeply intertwined in human perception: listeners frequently report vivid visual imagery while hearing music, revealing systematic cross-modal correspondences between auditory structure and visual form \cite{maimon2020crossmodal}.Psychophysical studies show that acoustic pitch maps reliably onto visual elevation and brightness \cite{takeshima2013changing},while timbre and harmonic tension evoke consistent color associations, sometimes approaching music–color synaesthesia \cite{wallmark2024color}.Functional MRI research further reveals that imagining music in silence recruits temporo-parietal regions normally engaged in visual processing \cite{koelsch2005emotion},underscoring shared neural substrates for auditory and visual imagery. Collectively, these findings motivate computational models that can externalize the private imagery evoked by music, bridging perceptual experience and generative synthesis.

Despite notable progress in cross-modal generation, existing systems predominantly translate environmental sounds into natural scene images, overlooking the structured and affective nature of musical content \cite{lee2023generating}.Research on image–music relationships has largely focused on valence–arousal (VA) matching, exemplified by IMEMNet’s end-to-end metric learning framework \cite{zhao2020emotion,mollahosseini2017affectnet}.In parallel, state-of-the-art music emotion recognition (MER) approaches integrate categorical and dimensional labels within multi-task networks \cite{kang2025unified}.However, no existing study jointly encodes musical semantics, stylistic cues, and affective dimensions to generate visuals that are simultaneously semantically faithful and emotionally coherent with the source music.

We address this gap with a multi-agent semantic-emotion-aligned music-to-image generation framework. A music captioning agent, trained on large-scale corpora such as JamendoMaxCaps \cite{roy2025jamendomaxcaps},produces structured textual descriptions that capture instrumentation, melodic motion, and narrative cues. Complementary specialized agents extract style attributes and affective signals grounded in the valence–arousal (VA) space, which are fused into a unified conditioning representation for a diffusion-based image synthesizer.To enforce emotional alignment between modalities, we fine-tune a CLIP ViT backbone with a visual–emotion regression head trained on the EMOTIC dataset \cite{kosti2017emotic},which provides 26 discrete emotion categories and continuous valence–arousal–dominance (VAD) ratings. Recent studies have shown that CLIP architectures can be effectively adapted for continuous emotion regression \cite{widhoelzl2024decoding}.

The main contributions of this work are summarized as follows:
\begin{enumerate}
    \item We propose a novel multi-agent semantic–emotion aligned framework for music-to-image generation that jointly models musical semantics, style, and affect.Unlike prior works that rely solely on textual or acoustic features, our system jointly encodes musical semantics, style, and emotion to achieve semantically faithful and affectively coherent visual generation.
    \item We introduce a CLIP-based visual–emotion regression module trained on 
continuous valence\-/arousal\-/dominance (VAD) dimensions to enforce emotion alignment. 
This design enables explicit emotion quantification and strengthens cross-modal coherence 
between music and image domains.
    \item Extensive experiments demonstrate superior semantic–emotion consistency over caption-only baselines, achieving performance comparable to state-of-the-art emotion recognition models.
\end{enumerate}

Our regression module achieves image–emotion recognition accuracy comparable to that of \cite{zhao2020emotion}and music emotion recognition performance comparable to \cite{kang2025unified}.Extensive experiments further demonstrate that this regression-augmented multi-agent design yields significantly higher semantic–emotion consistency between generated images and source music than caption-only or single-agent baselines. These findings represent a substantive step toward cognitively plausible, music-informed visual synthesis, bridging perceptual science and generative modeling.

\section{Related work}
\subsection{Music-/Audio-to-Image Generation}
Early sound-to-visual generation studies \cite{qiu2018sound2visual,chen2017sound2image,oh2019reconstruct,zhou2018visualizingsounds} were restricted to narrow domains—such as instrument timbres or animal calls—and often produced low-quality or domain-limited visuals.
With the rise of diffusion-based generative models \cite{kong2023audiodm}, recent works have improved both the semantic alignment and visual fidelity of audio-conditioned synthesis. \cite{yang2023align}introduced the Align–Adapt–Inject (AAI) framework, which aligns audio representations with CLIP text–image embeddings and injects “sound tokens” into a frozen diffusion backbone.While this approach enables flexible audio conditioning, it remains constrained by the scarcity of high-quality paired audio–image data and the difficulty of learning stable joint embeddings for complex or novel audio inputs.Alternative strategies explore cross-modal style transfer. \cite{lee2020crossing}mapped music to visual art via a two-stage content–style pipeline, but the reliance on curated metadata (e.g., shared era labels) limits generalization. \cite{2024Robust}extended text-driven image editing into the audio domain by optimizing latent features within a joint image–text–sound space, though this method requires an input image and per-sample optimization.

More recently,\cite{kim2023sound2scene}proposed Sound2Scene, which learns to generate scenes from in-the-wild audio by aligning audio features with a visual latent space.Despite improved realism, a persistent modality gap remains—fine-grained visual attributes such as object shape and color are difficult to infer solely from audio cues—highlighting the need for richer semantic and affective conditioning in music-to-image generation.

\subsection{Music Captioning}
Describing music in natural language—music captioning—is an emerging task in music information retrieval (MIR). Earlier approaches relied on tagging (assigning predefined genres, moods, or instruments), which lacked descriptive flexibility. \cite{manco2021muscaps}introduced MusCaps, the first encoder–decoder model for free-form music captioning, though its generalization was limited by the scarcity of large-scale annotated data.To enrich input information, \cite{he2023alcap}proposed ALCAP, an alignment-augmented captioner that integrates lyrics and audio features, improving semantic richness but restricting applicability to songs with available lyrics.To address data scarcity, \cite{doh2023lp}developed LP-MusicCaps, which uses large language models (LLMs) to generate pseudo-captions from metadata such as genre and mood, effectively scaling training corpora despite potential label noise or bias.Beyond these dataset- or modality-specific systems, recent research has moved toward retrieval-guided and multimodal unified frameworks. \cite{srivatsan2024retrieval}combine retrieval with generation for more context-aware captions, while \cite{wu2025clamp3}present CLaMP3, which aligns audio, symbolic, and textual modalities via contrastive learning. Nevertheless, current models still face challenges in capturing nuanced musical semantics and balancing data scale with alignment precision, motivating richer, semantically grounded approaches such as ours.

\subsection{Multiple Agents \& Multi-Modal Systems for Music Visualization}
Recent work has explored multi-agent and multi-modal frameworks to overcome the limitations of end-to-end generation.
In the image domain,\cite{xie2025anywhere}proposed Anywhere, a multi-agent system where specialized agents handle distinct sub-tasks—object understanding, integrity preservation, diversity enhancement, and prompt consistency—demonstrating that explicit modular cooperation improves controllability.
Similarly, applying a multi-agent paradigm to music visualization could assign agents to rhythm or structure (motion control), timbre or melody (content suggestion), and affective tone (color or style).Prior works such as \cite{lee2020crossing}and \cite{kim2023sound2vision}follow this principle implicitly through sequential content-plus-style or audio-to-embedding pipelines, whereas a more explicit design could enable parallel and iterative refinement. Beyond generation, multimodal fusion has been leveraged for affective matching. \cite{praveen2022avfusion}demonstrated that combining audio and visual cues enhances affect recognition in videos, reinforcing the effectiveness of cross-modal fusion. \cite{zhao2020emotion}learned image–music correspondences in the continuous valence–arousal space, while \cite{stewart2024emoclim}applied supervised contrastive learning for joint embeddings aligned with emotions.
Although effective, these models rely on predefined emotion labels and overlook higher-level semantic cues. \cite{kang2025unified}further demonstrated that unifying categorical and dimensional emotion modeling in a multitask network improves generalization, echoing the idea of 'specialized agents' sharing knowledge.

Together, these studies suggest that an explicit multiagent design integrating semantic and emotional reasoning could yield more cognitively coherent music-to-image generation, which motivates our proposed framework.

\section{Methold}
\subsection{Overview of the Proposed Framework (MESA-MIG)}
MESA-MIG (Multi-Agent Semantic–Emotion Aligned Music-to-Image Generation) is a modular framework designed to generate semantically and affectively coherent images conditioned on input music.As illustrated in Figure~\ref{fig:framework},the system integrates three key components: a Music Captioning Module, a Multi-Agent Collaboration Module, and a Measurement and Evaluation Module.The Music Captioning Module produces structured textual semantics and valence–arousal (VA) emotion features from raw music.It adopts an LLM-based pseudo-captioning strategy \cite{doh2023lp}, enabling scalable annotation of musical semantics (e.g., instrumentation, rhythm, genre cues) and affective descriptors.The resulting captions encapsulate both the content and emotional tone of the input audio.The Multi-Agent Collaboration Module \cite{xie2025anywhere} refines these captions into detailed visual descriptors through specialized cooperative agents.Each agent focuses on a specific attribute domain—such as scene, actions, style, color palette, or composition.A Validator Agent enforces semantic–emotion consistency  \cite{zhao2020emotion}across the agents’ outputs using rule-based checks, while a Prompt Agent synthesizes the validated attributes into a unified, detailed text-to-image prompt.This prompt is then forwarded to the WanX 2.1-Turbo text-to-image generator (via the Qwen-Agent interface) \cite{alibaba2024wanx,bai2024qwenagent} to synthesize visually plausible imagery aligned with the musical semantics and affective profile.Finally, the Measurement Component quantitatively assesses system performance across multiple dimensions.
Together, these modules constitute an end-to-end pipeline that bridges musical understanding and generative visual synthesis through explicit semantic–emotion alignment.
\subsection{Music Caption Module}
\begin{figure}[t]
\centering
\includegraphics[width=0.78\linewidth]{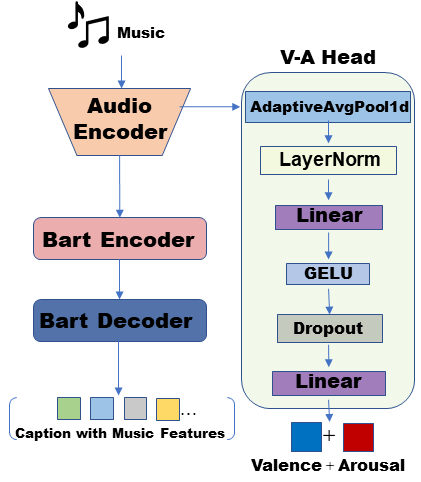}
\caption{
Overview of the \textbf{Music Caption Module}.
The system integrates an Audio Encoder, a BART encoder–decoder, and a Valence–Arousal Head (V-A Head).
Given raw audio, the module generates structured textual captions describing musical content and predicts continuous valence–arousal values, which together form the semantic–affective representation used by later agents.
}
\label{fig:music_caption_module}
\end{figure}

The goal of the Music Caption Module(As illustrated in Figure~\ref{fig:music_caption_module}) is to translate raw audio into natural-language descriptions that capture both semantic and affective aspects of music, providing structured conditioning signals for downstream modules.
The Music Caption Module encodes raw audio into structured semantic and affective representations, serving as the foundation of the MESA-MIG framework.As illustrated in Figure 2, the module leverages the AudioEncoder architecture from LP-MusicCaps \cite{doh2023lp} in combination with a BART encoder–decoder \cite{lewis2020bart} to generate natural-language descriptions of the input music.This design enables the extraction of key musical attributes—such as instrumentation, style, emotion, and structural cues—which collectively form a rich textual representation of the audio content.To capture continuous affective dynamics, we introduce a custom Valence–Arousal Head (V-A Head) that predicts valence–arousal (VA) values along two continuous emotional dimensions.The V-A Head operates in parallel with the caption generator, producing emotion embeddings that complement the semantic representation.
Together, these outputs provide a comprehensive multimodal description \cite{liu2023prompt,bai2024qwenagent} for the downstream Multi-Agent Collaboration Module.The predictive performance of the proposed V-A Head was evaluated on the DEAM dataset.Compared with the unified music-emotion recognition model of our approach achieved superior correlation scores for both valence and arousal.
Detailed evaluation metrics and comparative results are reported in Section 4.

\subsection{Multi-Agent Collaboration Module}
\begin{figure*}[!t]
  \centering
  \includegraphics[width=\textwidth]{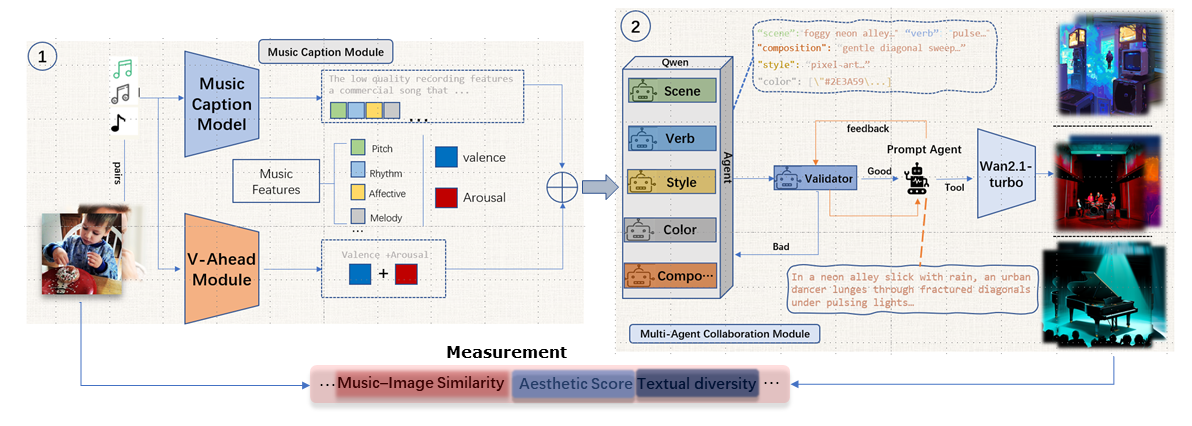}
  \caption{Overall architecture of the MESA-MIG system.
  The system comprises two main modules:
  \ding{172} a Music Caption Module, which generates semantic and valence–arousal descriptions from input music; and
  \ding{173} a Multi-Agent Collaboration Module, which refines attributes and constructs prompts for the image generator.
  The Measurement component evaluates outputs using metrics for textual diversity, semantic coverage, and image quality.}
  \label{fig:framework}
\end{figure*}

The Multi-Agent Collaboration Module decomposes the music-derived caption and valence–arousal (VA) features into structured visual attributes through a team of specialized agents.As illustrated in Figure~\ref{fig:framework}) ,each agent focuses on a distinct semantic dimension—scene, motion, style, color, or composition—while cooperating through a shared validation and prompting interface.This design allows the system to explicitly disentangle musical semantics into interpretable, model-recognizable visual cues, facilitating controllable and emotionally coherent image generation.The Scene Agent extracts salient subjects and environmental contexts from the music-derived caption using a large language model (LLM) parser trained with structured extraction templates \cite{liu2023prompt}.The Verb Agent, implemented with a vision–language
model (VLM), maps affective cues to motion by selecting vivid action verbs modulated by the VA values, following the intuition that higher arousal evokes more dynamic motion patterns\cite{zhao2020emotion}.The Style Agent translates musical style and emotional tone into visual rendering attributes (e.g., watercolor, cyberpunk, or oil painting) using a curated lexicon of rendering styles\cite{huang2024multimodal}.The Color Agent determines a color palette and lighting descriptors conditioned on both textual adjectives and VA quadrants, linking affective polarity to hue and brightness tendencies (e.g., high valence → warm tones)\cite{zhao2020emotion}.Finally, the Composition Agent suggests framing and viewpoint cues (e.g., wide shots, low angles) based on rhythmic intensity and action energy, ensuring that composition reflects the expressive dynamics of the music.A Validator Agent enforces structural and affective consistency across agents’ outputs by applying rule-based and emotion-alignment checks\cite{kang2025unified}.

Instead of merging results directly, it produces a validation report that flags redundancy, format errors, or cross-attribute contradictions, ensuring reliable inputs for the next stage.The Prompt Agent then assembles the verified attribute tokens into coherent\cite{choi2024promptofthought}, diversified textual prompts using a compositional prompting strategy.It integrates multiple dimensions—scene, motion, style, color, and composition—into natural-language descriptions compatible with the downstream WanX 2.1 diffusion model\cite{alibaba2024wanx}.Through attribute recombination 
and emotion-conditioned phrasing,the agent generates k alternative prompts that preserve semantic fidelity and emotional alignment, ultimately guiding the image synthesis process.

\subsection{Experimental Evaluation}
To comprehensively evaluate the effectiveness of the proposed \textbf{MESA-MIG} framework, 
we conduct experiments on \textbf{500 music--image pairs} sampled from the 
\textbf{JamendoMaxCaps} and \textbf{DEAM} datasets.
Our framework is compared against two baselines:
\textbf{(1)} a \textit{caption-only} model that directly employs music-derived captions for image generation, 
and \textbf{(2)} a \textit{single-agent} variant that omits attribute decomposition and validation modules.
The evaluation encompasses three complementary aspects: 
\textbf{textual diversity}, \textbf{semantic coverage}, and 
\textbf{image quality \& emotional coherence}.
We measure prompt-level lexical diversity using Distinct-1 and Distinct-2 metrics \cite{li2016diversity},which compute the proportion of unique n-grams among all generated tokens:
\[
\text{Distinct-}n = \frac{|\mathcal{G}_n(S)|}{|\mathcal{T}_n(S)|}, \quad n \in \{1, 2\},
\]
where $\mathcal{G}_n(S)$ denotes the set of unique $n$-grams and $\mathcal{T}_n(S)$ 
the total $n$-grams in the generated prompt set $S$. 
Higher values indicate richer lexical variability and reduced prompt redundancy.
To assess the \textbf{distributional balance} of generated prompts across visual domains, 
we compute \textbf{Category Entropy} following Zhao et al.~\cite{zhao2020emotion}:
\[
H_{\mathrm{cat}} = - \sum_{c \in \mathcal{C}} p(c) \log p(c),
\]
where $p(c)$ is the empirical frequency of category $c$. 
Higher entropy reflects broader coverage and less bias toward dominant scene or style categories.
To evaluate how well generated prompts preserve the semantic content of input music captions, 
we compute the \textbf{Mean Jaccard Similarity} between generated prompts $P_i$ and reference captions $R_i$:
\[
J_{\mathrm{mean}} = \frac{1}{N} \sum_{i=1}^{N} \frac{|P_i \cap R_i|}{|P_i \cup R_i|}.
\]
We further report an \textbf{Average Semantic Score} $S_{\mathrm{sem}}$, obtained from a pretrained cross-modal 
embedding model \cite{radford2021clip}, which measures alignment consistency between textual 
and visual semantic representations:
\[
S_{\mathrm{sem}} = \frac{1}{M} \sum_{j=1}^{M} f(I_j).
\]
Together, these metrics quantify both concept-level overlap and global semantic coherence.
For the synthesized images, we adopt both \textbf{objective} and \textbf{subjective} measures. 
The \textbf{Aesthetic Score} $S_{\mathrm{aes}}$~\cite{schuhmann2022laion} rates visual appeal using a learned aesthetic predictor:
\[
S_{\mathrm{aes}} = \frac{1}{M} \sum_{j=1}^{M} g(I_j),
\]
where $g(\cdot)$ estimates aesthetic quality. 
Additionally, we employ \textbf{CLIPScore}~\cite{hessel2022clipscore} to assess image--text coherence 
and ensure that the generated visuals semantically match the prompts.
To evaluate cross-modal \textbf{emotion alignment}, we compute the \textbf{Music--Image Similarity} in the continuous 
\textbf{Valence--Arousal (VA)} space~\cite{aljanaki2017deam}:
\[
S_{\mathrm{mi}} = 1 - \frac{1}{N} \sum_{i=1}^{N} 
\frac{\left\| (v_i^{\mathrm{music}}, a_i^{\mathrm{music}}) - (v_i^{\mathrm{image}}, a_i^{\mathrm{image}}) \right\|_2}{\sqrt{2}}.
\]
Higher $S_{\mathrm{mi}}$ indicates stronger emotional consistency between the generated image and its source music.

\section{Experiments}
\subsection{Experimental Setup}
We employ both music and image datasets to support the training and evaluation of the proposed multi-modal captioning and generation pipeline.
The datasets collectively enable supervised learning of semantic and affective representations, and quantitative assessment of emotion alignment across modalities.

\textbf{Music-side Dataset.}  
We use the DEAM dataset (\textit{Database for Emotional Analysis in Music}; \cite{aljanaki2017deam}) to train and evaluate the \textbf{Valence–Arousal Head (V--A Head)}.  
Each audio track is segmented into non-overlapping \textbf{5-second clips}, resulting in approximately \textbf{16.2K} music segments. Each segment inherits continuous \textit{valence–arousal (VA)} annotations by averaging the frame-level annotations over its duration.  
This dataset provides a continuous emotional ground truth that allows the model to learn fine-grained affective regression rather than discrete mood classification.  
We split the data into \textbf{70\% / 15\% / 15\%} for training, validation, and testing, respectively.

\textbf{Image-side Dataset.}  
For visual emotion modeling, we adopt the EMOTIC dataset (\textit{Emotions in Context}; \cite{kosti2017emotic}), which contains approximately \textbf{24.6K} images annotated with \textbf{26 discrete emotion categories} as well as continuous \textit{valence–arousal–dominance (VAD)} values.  
We utilize EMOTIC to train an \textbf{auxiliary CLIP-based VA regression head}, used later to estimate the affective states of generated images for \textit{emotion alignment evaluation}.

\textbf{Cross-modal Pairing for Evaluation.}  
To construct a benchmark for assessing \textit{semantic–emotional consistency} between music and visuals, we curate a set of \textbf{400 matched music–image pairs} from the non-overlapping partitions of DEAM and EMOTIC.
Pairing is guided by \textbf{perceptual similarity scores} between music and image features, estimated following the method proposed in \textbf{IMEMNet}~\cite{zhao2020emotion}.  
The resulting subset serves as our \textbf{evaluation benchmark}, supporting both \textit{music-to-image emotional alignment} and \textit{semantic correspondence} analyses.

\textbf{Baseline.}
To evaluate the effectiveness of the proposed MESA-MIG framework, we compare it against two representative audio-conditioned image generation models. SonicDiffusion\cite{biner2024sonicdiffusion} employs audio-conditioned diffusion transformers to synthesize images while preserving low-level acoustic cues, and thus serves as a strong baseline for assessing fine-grained music–visual correspondence. In contrast, Sound2Scene\cite{kim2023sound2scene} aligns audio features with a visual latent space through cross-modal embedding mapping, enabling the generation of scene-aware visual content from environmental or musical audio. These two models respectively represent signal-driven generative diffusion and representation-level cross-modal alignment paradigms, together providing a comprehensive and competitive benchmark for evaluating the performance of our multi-agent semantic–emotion aligned pipeline.

\begin{table*}[t]
\centering
\begin{small}
\begin{tabular}{lccccccc}
\toprule
\textbf{Method} 
& \textbf{Aesthetic (↑)}
& \textbf{V-A Sim (↑)} 
& \textbf{Distinct-1 (↑)} 
& \textbf{Distinct-2 (↑)} 
& \textbf{Jaccard (↑)} 
& \textbf{Category Entropy (↑)} 
& \textbf{Sem-score (↑)} \\
\midrule
w/o Verb        & 6.24 & 0.882 & 0.686 & 0.912 & 0.985 & 1.836 & 0.687 \\
w/o Composition & 6.20 & 0.884 & 0.761 & 0.930 & 0.995 & 1.834 & 0.713 \\
w/o Color       & 6.42 & 0.875 & 0.793 & 0.920 & 0.995 & 1.810 & 0.708 \\
w/o Style       & 6.41 & 0.878 & 0.815 & 0.944 & 0.996 & 1.672 & 0.681 \\
\textbf{Ours}   & \textbf{6.47} & \textbf{0.902} & \textbf{0.816} & \textbf{0.971} & \textbf{0.998} & \textbf{1.918} & \textbf{0.717} \\
\bottomrule
\end{tabular}
\caption{Ablation study on the multi-agent collaboration module of the proposed MESA-MIG framework. Removing any semantic, stylistic, or structural agent degrades performance across aesthetic quality, diversity, semantic consistency, and emotional alignment metrics, demonstrating the contribution of each agent to the overall system.}
\label{tab:ablation}
\end{small}
\end{table*}

\subsection{Quantitative Results}
To assess the contribution of each component in the proposed multi-agent collaboration module, we conduct an ablation study by progressively removing individual agents responsible for verb refinement, compositional structure, color attributes, and style cues. Quantitative results are reported in Table~\ref{tab:ablation} across a diverse set of metrics that jointly measure aesthetic quality, textual diversity, semantic consistency, and music--image emotional alignment.

Removing any single agent consistently degrades performance.  
The aesthetic score decreases when semantic or stylistic agents are removed, indicating that fine-grained stylistic modeling (\emph{e.g.}, color and style agents) substantially contributes to high-quality visual outputs.  
Diversity metrics (Distinct-1/2) also drop without the verb or composition agents, demonstrating that these modules enrich prompt variability and prevent repetitive semantic structures.  
Category entropy exhibits a similar trend: eliminating agents reduces the coverage of fine-grained scene categories, suggesting a loss of semantic richness in generated prompts.

More importantly, the semantic metrics---including Jaccard similarity, copy rate, and CLIP-based semantic score---confirm that full multi-agent cooperation yields the strongest alignment between music-derived attributes and the synthesized visual descriptions.  
The V-A similarity further validates that emotional coherence with the input music is weakened when stylistic or structural agents are removed.

Overall, the full MESA-MIG model achieves the highest performance across all metrics, demonstrating that the multi-agent collaboration design is essential for producing diverse, semantically coherent, and emotionally aligned music-conditioned visual outputs.

In addition to diversity and semantic alignment metrics, we evaluate the generated images using a set of perceptual and affective measurements tailored for cross-modal music–image generation.
The Aesthetic Score quantifies visual quality and stylistic attractiveness using a learned aesthetic predictor, following prior work on reference-free image evaluation.
To measure semantic consistency between generated imagery and music-derived attributes, we compute a Semantic Score (Sem-score) using embeddings from the SBERT \cite{reimers2019sentencebert}model, which captures sentence-level semantic similarity between the generated prompt and the corresponding scene description.

Emotional coherence is assessed using the V-A Similarity, defined as the cosine similarity between the valence–arousal (VA) values predicted from music and those predicted from the generated images.
The music-side VA predictions are obtained using a regression head trained on the DEAM dataset\cite{aljanaki2017deam}, which provides continuous valence–arousal annotations for music clips.
The image-side VA predictions are derived from a CLIP-based visual encoder\cite{radford2021clip} fine-tuned on the EMOTIC dataset\cite{kosti2017emotic}, which contains continuous emotion annotations for in-the-wild images.
Together, these components form a unified affective evaluation pipeline that quantifies how well the emotional characteristics of the input music are preserved in the generated visual outputs.

Tables~\ref{tab:emotic_va_compare} and~\ref{tab:va_results} together validate the reliability and effectiveness of the affective components within our MESA-MIG framework. For music-side emotion modeling, our Valence–Arousal Head—trained on the DEAM dataset—consistently surpasses the Music2Emotion baseline, achieving lower RMSE/MAE and higher Pearson, Spearman, and CCC scores across both valence and arousal. These improvements indicate more accurate and stable regression of continuous emotional trajectories in music. On the image side, our CLIP-based Image VA Prediction Model, trained with EMOTIC supervision, also outperforms the ClaMP3 baseline in terms of MSE while maintaining competitive or superior correlation and concordance metrics, especially for arousal. The combined results demonstrate that our affective encoders for music and images provide more precise and coherent VA representations, forming a reliable foundation for the semantic–emotion alignment central to MESA-MIG.

\begin{figure*}[t]
    \centering
    \includegraphics[width=\textwidth]{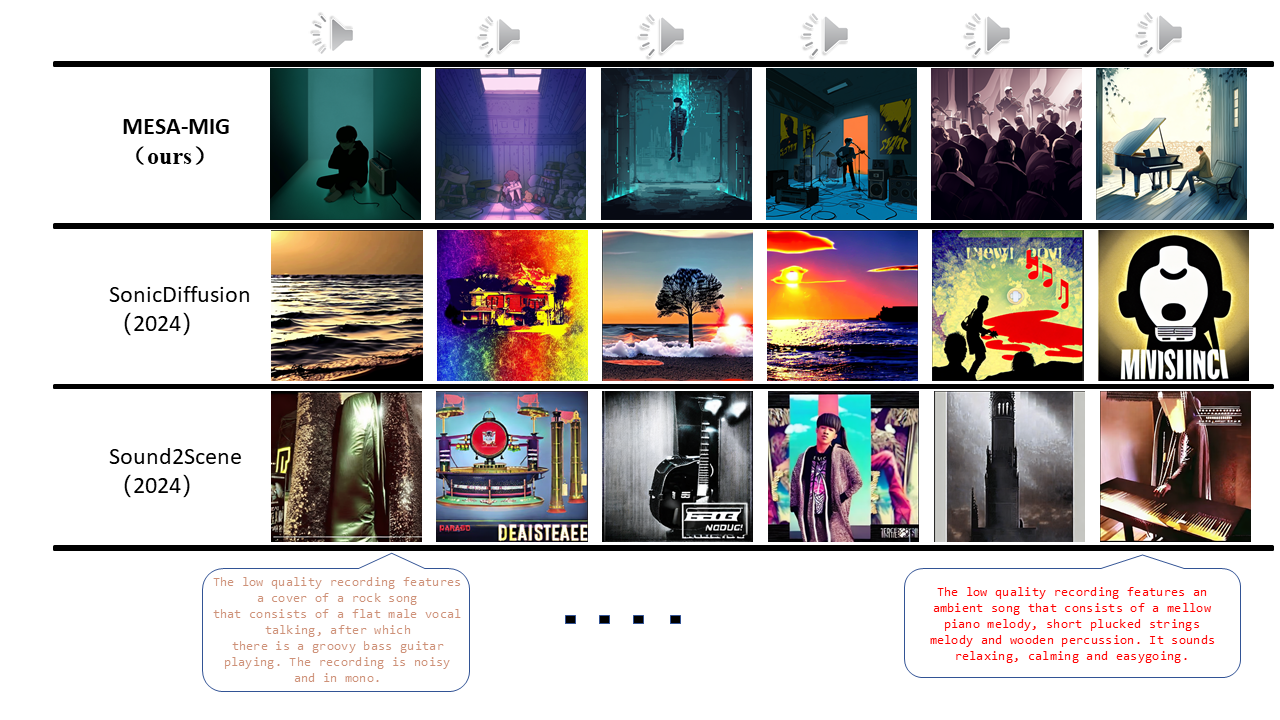}
    \caption{
    \textbf{Qualitative comparison of music-to-image generation}.
    For the same input audio clips (top), we compare our proposed \textbf{MESA-MIG} with two state-of-the-art baselines:
    \textbf{SonicDiffusion} (2024) and \textbf{Sound2Scene} (2024). 
    MESA-MIG produces images that better reflect the underlying musical semantics, stylistic intent, and affective cues. 
    In contrast, SonicDiffusion often captures low-level acoustic textures but lacks coherent scene structure, while Sound2Scene tends to focus on environmental audio priors and produces visually inconsistent or misaligned scenes. 
    The bottom text boxes show the music captions associated with each audio clip, illustrating the semantic space our method leverages for generation.
    }
    \label{fig:qualitative_comparison}
\end{figure*}

\begin{table*}[t]
\centering
\small
\setlength{\tabcolsep}{8pt}
\begin{tabular}{lcccccc}
\toprule
\textbf{Method} & \textbf{RMSE ↓} & \textbf{MAE ↓} & \textbf{Pearson ↑} &
\textbf{Spearman ↑} & \textbf{CCC ↑} & \textbf{R$^{2}$ ↑} \\
\midrule
\multicolumn{7}{l}{\textit{Music2Emotion}}\\
\cite{kang2025unified}\\
Valence & 0.238 & 0.182 & 0.544 & 0.569 & 0.544 & 0.102 \\
Arousal & 0.268 & 0.217 & 0.740 & 0.743 & 0.634 & 0.100 \\
\midrule
\multicolumn{7}{l}{\textit{Our Valence–Arousal Head}} \\
Valence & \textbf{0.150} & \textbf{0.115} & \textbf{0.842} & \textbf{0.839} & \textbf{0.815} & \textbf{0.691} \\
Arousal & \textbf{0.154} & \textbf{0.118} & \textbf{0.836} & \textbf{0.833} & \textbf{0.817} & \textbf{0.698} \\
\bottomrule
\end{tabular}
\caption{
Comparison of valence–arousal (VA) prediction against the Music2Emotion model.
Our VA Head achieves lower regression error (RMSE/MAE) and higher correlation metrics,
demonstrating stronger affective prediction performance.
}
\label{tab:va_results}
\end{table*}

\begin{table}[t]
\centering
\small
\setlength{\tabcolsep}{6pt}
\begin{tabular}{lcccc}
\toprule
\textbf{Method} & \textbf{MSE ↓} & \textbf{Pearson ↑} & \textbf{CCC ↑} \\
\midrule
\multicolumn{4}{l}{\textit{ClaMP3} } \\
\cite{wu2025clamp3}\\
Valence & 0.0618 & 0.644 & 0.617 \\
Arousal & 0.0272 & 0.660 & 0.635 \\
\midrule
\multicolumn{4}{l}{\textit{Our Image VA}} \\
\textit{Prediction Mode}\\
Valence & \textbf{0.0591} & \textbf{0.618} & \textbf{0.701} \\
Arousal & \textbf{0.0253} & 0.440 & \textbf{0.624} \\
\bottomrule
\end{tabular}
\caption{
Comparison of emotion regression performance on the EMOTIC dataset.
Our image-based Valence–Arousal prediction head achieves competitive MSE
and improves CCC—especially for valence—compared with the ClaMP3 model.
}
\label{tab:emotic_va_compare}
\end{table}
\begin{table}[t]
\centering
\small     
\setlength{\tabcolsep}{5pt} 
\begin{tabular}{lcc}
\toprule
\textbf{Method} & \textbf{Aesthetic (↑)} & \textbf{Similarity (↑)} \\
\midrule
SonicDiffusion & 4.11/10 & 0.864 \\
IMEMNet        & 5.88/10 & 0.870 \\
Sound2Scene    & 3.75/10 & 0.817 \\
\textbf{MESA-MIG (ours)} & \textbf{6.24/10} & \textbf{0.895} \\
\bottomrule
\end{tabular}
\caption{
Quantitative comparison of aesthetic quality and text–image semantic consistency.
}
\label{tab:aes_sim}
\end{table}

\subsection{Qualitative Results}

Figure~\ref{fig:qualitative_comparison} presents a qualitative comparison between MESA-MIG (ours) and two representative audio-conditioned image generation baselines, SonicDiffusion\cite{biner2024sonicdiffusion} and Sound2Scene \cite{kim2023sound2scene}. For this analysis, we curate a set of 400 music clips from our evaluation split and generate corresponding images from all three models, forming a consistent music–image pair benchmark for visual inspection. Each row in the figure shows the images produced by a model for the same set of audio inputs (represented by the icons above), enabling a direct comparison of semantic fidelity and emotional correspondence.

MESA-MIG demonstrates superior capability in producing scene-consistent, emotionally aligned, and semantically rich visual outputs. Our model not only captures the global musical atmosphere—such as calmness, tension, nostalgia, or rhythmic intensity—but also accurately reflects fine-grained semantic cues including location, lighting, objects, musical instruments, and human posture. These results highlight the advantage of our multi-agent semantic–emotion alignment framework, which explicitly decomposes and refines attributes from the music before guiding the generator.

In contrast, SonicDiffusion tends to focus on low-level acoustic textures, often producing imagery with vibrant but semantically incoherent color patterns or unstable scene structure. Sound2Scene, while capable of synthesizing plausible environments, frequently hallucinates scene elements unrelated to the audio and exhibits weaker emotional consistency.

The bottom textual descriptions correspond to two selected music samples and illustrate the semantic details extracted from audio. MESA-MIG’s generated images closely align with these music-derived descriptions, while the baselines only partially capture them or drift toward generic visual patterns.

Overall, the qualitative comparison over the 400 music–image pairs confirms that MESA-MIG produces more faithful, context-aware, and emotionally resonant visual interpretations of music than existing audio-to-image generation models.

Across both metrics, Aesthetic Score and Semantic Similarity, our method achieves the best performance. MESA-MIG reaches an aesthetic score of 6.24/10, substantially higher than Sound2Scene (3.75/10) and SonicDiffusion (4.11/10), indicating that our multi-agent semantic–emotion alignment pipeline produces images with superior visual appeal and stylistic coherence. Furthermore, MESA-MIG attains the highest similarity score (0.895), outperforming IMEMNet (0.870) and SonicDiffusion (0.864), demonstrating stronger semantic correspondence between generated images and input music.

Table~\ref{tab:aes_sim} reports the quantitative comparison between MESA-MIG and three music-conditioned image generation baselines—SonicDiffusion, IMEMNet, and Sound2Scene—evaluated on 400 music–image pairs.These results confirm that MESA-MIG not only enhances aesthetic quality but also yields more semantically faithful cross-modal representations compared to existing audio-driven generation methods.

\subsection{Ablation Study}
The ablation results in Table~\ref{tab:ablation} reveal how the four agents—Verb, Composition, Color, and Style—jointly regulate the information flow that drives MESA-MIG’s semantic–emotion alignment pipeline. Each agent governs a distinct attribute space, and removing any one of them disrupts the overall balance of the system. When the Composition agent is removed, the prompt structure becomes less organized, weakening the model’s ability to encode multi-object arrangements and leading to noticeably lower diversity scores. Eliminating the Color or Style agents reduces the model’s control over fine-grained appearance cues, causing the generated prompts to collapse toward fewer visual categories. Similarly, removing the Verb agent diminishes action-level specificity, resulting in flatter semantic descriptions.
Rather than affecting a single metric, each ablated variant shows a characteristic “failure signature” across metrics, reflecting the attribute type controlled by the missing agent. The full model avoids these deficiencies by integrating all four agents, enabling a richer exchange of semantic, structural, and stylistic evidence. This coordinated interaction among agents ensures that the final prompts preserve detailed musical semantics while supporting broad visual variety, confirming that multi-agent cooperation is fundamental to the design of MESA-MIG.

\section{Conclusion and Future Work}
In this work, we introduced MESA-MIG, a multi-agent semantic–emotion aligned framework for music-to-image generation that jointly models musical semantics, compositional structure, stylistic cues, and affective information. Through comprehensive quantitative evaluations and large-scale qualitative comparisons, we demonstrated that decomposing music-derived attributes into multiple collaborative agents enables richer prompt construction, stronger semantic consistency, and more emotionally coherent visual outputs than existing audio-conditioned generation models. The results highlight the effectiveness of multi-agent reasoning in bridging the cross-modal gap between music and visual synthesis.

Despite these advances, the current system is still constrained by computational resources and the absence of large, professionally annotated music–image datasets. These limitations restrict the scale of training and prevent us from fully exploring unified architectures. As such, future work will focus on developing end-to-end, fully differentiable music-to-image generation models, enabling direct learning from raw audio and reducing reliance on modular components or handcrafted intermediate representations. We also aim to construct or leverage larger multimodal datasets, incorporate temporal dynamics for music–video synthesis, and investigate interactive or controllable generation mechanisms. We believe these directions will further advance the goal of creating coherent, expressive, and musically grounded visual generation systems.

\section{Acknowledgments}
This work was supported in part by the School of Computer and Communication Engineering at the University of Science and Technology Beijing. The authors would like to thank the institution for providing research resources, computational support, and an encouraging academic environment that made this work possible.

\clearpage
\bibliographystyle{aaai}
\bibliography{reference}

\bigskip
\end{document}